# Modeling anisotropic charged neutron star in isotropic coordinates

Ksh. Newton Singh[1], Narendra Pradhan[1], Manuel Malaver[2]

[1]Department of Physics, National Defence Academy, Khadakwasla, Pune-411023, India
[2]Department of Basic Sciences, Maritime University of the Caribbean, Catia la Mar, Venezuela

**Email address:**
ntnphy@gmail.com (K. N. Singh), npradhan20569@gmail.com (N. Pradhan), mmf.umc@gmail.com (M. Malaver)

**Abstract:** We present a spherically symmetric solution of the general relativistic field equations in isotropic coordinates for charged fluid with pressure anisotropy, compatible with a super dense star modeling. Further, we have constructed an anisotropic model of super dense star with all degree of suitability. We also observed that by increasing anisotropy, the maximum mass of super dense stars also decreases.

## 1. Introduction

Since neutron stars were discovered, many different properties are being observed and developing new models to explain its phenomenological behaviors. Neutron stars are form when massive stars of masses more than $8M_\odot$ collapse by its own gravity end with a highly energetic explosion called the "supernova" leaving behind a super-dense core mostly composed of neutrons [1]. Neutron stars are famous for its immense surface magnetic field which is of the order of $10^{12} - 10^{14} G$. Some of the theories of structure of neutron stars also suggest the existence of *super-fluid neutrons* and *superconducting protons*. The *glitches* in pulsars support the existence of super-fluid neutrons.

After the formulation of Einstein-Maxwell field equations, the relativists have been proposing different models of immense gravity astrophysical objects by considering the distinct nature of matter or radiation (energy-momentum tensor) present in them. Such models successfully explain the characteristics of massive objects like quasar, neutron star, pulsar, quark star, black-hole or other super-dense object.

It is well known that the presence of some charge may avert the gravitational collapse by counter balancing the gravitational attraction by the electric repulsion in addition to the pressure gradient. Ivanov [2] proposed a model for charged perfect fluid and concluded that the inclusion of charge inhibits the growth of space time curvature which has a great role to avoid singularities. Bonnor [3] pointed out that a dust distribution of arbitrarily large mass and small radius can remain in equilibrium against the pull of gravity by a repulsive force produced by a small amount of charge. Thus it is desirable to study the implications of Einstein –Maxwell field equations with reference to the general relativistic prediction of gravitational collapse. For this purpose charged fluid ball models are required. The external field of such ball is to be matched with *Reissner –Nordstrom* solution.

In the formulism of realistic model of super dense stars, it is also important to include the pressure anisotropy. Bowers and Liang [4] extensively discuss the effect of pressure anisotropy in general relativity. At a density of the order of $10^{15}$ $g/cm^3$, nuclear matter may be anisotropic when its interactions need to be treated relativistically [5]. If there is existence of solid stellar core or presence of a type-3A super-fluid, one can also include pressure anisotropy [6]. Sokolov [7] suggest that a different kind of phase transition may lead to pressure anisotropy or pion condensation can generate anisotropy [8]. Since the neutron star has immense magnetic field, this also generate anisotropic pressure inside a compact star [9]. Usov [10] suggest that strong electric field may also cause pressure anisotropy.

Das et al. [11], Pant et al. [12,13], Pant and Negi [14], Kiess [15], Maurya and Gupta [16,17], Pant [18, 19], Murad and Fatima [20] and Malaver [21,22] obtained the class of well behave exact solutions with pressure isotropic . These solutions are well behaved and describe the interior of the super-dense astrophysical object with charged matter however anisotropy is not concerned.

In many papers, pressure anisotropy has been discussed in canonical coordinate [23,24,25,26,27], although the effects of anisotropy is never being accounted. In an article by Pant et al. [28] include the anisotropy term and solved for well behaved solution.

In this paper we introduce pressure anisotropy of a charge fluid ball in isotropic coordinate. Our object is to find a new exact solution of Einstein-Maxwell field equation by introducing anisotropy in pressure which will help in describing more realistic structure of super dense stars. In our analysis we assume a solution of $e^{\omega/2}$ and solve for $e^{\nu/2}$. In solving the master equation we assume a particular form of charge $q$ and pressure anisotropy factor $\Delta$ in such a way that the differential equation reduce to Riccati equation. The interior solution should be continuously matched with the exterior solution of *Reissner-Nordstrom*. To find a physically viable solution we choose the anisotropic pressure parameter $\alpha$, charge parameter $K$ and *Reissner-Nordstrom* parameter $d$ in such a way that it satisfies all the properties of a well behave solution mentioned in Sect. 2.

## 2. Conditions for Well Behaved Solution

For well behaved nature of the solutions for anisotropic fluid sphere should satisfy the following conditions [26,27,28]:

(i) The solution should be free from physical and geometrical singularities i.e. finite and positive values of central pressure, central density and non zero positive values of $e^{\omega}$ and $e^{\nu}$.

(ii) The solution should have positive and monotonically decreasing expressions for pressure and density ($\rho$ and $p$) with the increase of $r$. The solution should have positive value of ratio of pressure-density and less than 1 (weak energy condition) and less than 1/3 (strong energy condition) throughout within the star, monotonically decreasing as well.

(iii) The casuality condition should be obeyed i.e. velocity of sound should be less than that of light throughout the model. In addition to the above the velocity of sound should be decreasing towards the surface i.e. $\frac{d}{dr}\left(\frac{dp_r}{d\rho}\right)<0$ or $\left(\frac{d^2p_r}{d\rho^2}\right)>0$ and $\frac{d}{dr}\left(\frac{dp_\perp}{d\rho}\right)<0$ or $\left(\frac{d^2p_\perp}{d\rho^2}\right)>0$ for $0 \leq r \leq r_b$ i.e. the velocity of sound is increasing with the increase of density. In this context it is worth mentioning that the equation of state at ultra-high distribution has the property that the sound speed is decreasing outwards [29].

(iv) $\frac{p}{\rho} \leq \frac{dp}{d\rho}$ , everywhere within the ball.
$\gamma = \frac{d \log_e P}{d \log_e \rho} = \frac{\rho}{p}\frac{dp}{d\rho} \Rightarrow \frac{dp}{d\rho} = \gamma \frac{p}{\rho}$, for realistic matter $\gamma \geq 1$.

(v) The red shift $Z$ should be positive, finite and monotonically decreasing in nature with the increase of $r$.

(vi) Electric field intensity $E$, such that $E_{r=0} = 0$, is taken to be monotonically increasing.

(vii) The anisotropy factor $\Delta$ should be zero at the center and increasing towards the surface.

Under these conditions, we have to assume the one of the gravitational potential component in such a way that the field equation (8) can be integrated and solution should be well behaved. Further, the mass of such modeled super dense object can be maximized by assuming surface density, for Neutron star $\rho_b = 2 \times 10^{14}$ $g/cm^3$ [30].

## 3. Field Equations in Isotropic Coordinates

We consider the static and spherically symmetric metric in isotropic coordinates

$$ds^2 = -e^{\omega}\left[\,dr^2 + r^2(d\theta^2 + \sin^2\theta\, d\phi^2)\,\right] + c^2 e^{\nu} dt^2 \quad (1)$$

where $\omega$ and $\nu$ are functions of $r$.

The Einstein-Maxwell field equations of gravitation for a non empty space-time are

$$\begin{aligned}R^i_j - \frac{1}{2}R\delta^i_j &= -\frac{8\pi G}{c^4}T^i_j \\ &= -\frac{8\pi G}{c^4}[(p_\perp + \rho c^2)v^i v_j - p_\perp \delta^i_j + (p_r - p_\perp)\chi_j \chi^i \\ &\quad + \frac{1}{4\pi}(-F^{im}F_{jm} + \frac{1}{4}\delta^i_j F_{mn}F^{mn})] \end{aligned} \quad (2)$$

where $R_{ij}$ is Ricci tensor, $T_{ij}$ is energy-momentum tensor, $R$ is the scalar curvature and $F_{mn}$ is the electromagnetic field tensor, $p_r$ denotes the radial pressure, $p_\perp$ is the transversal pressure, $\rho$ the density distribution, $\chi^j$ is the unit space-like vector in the radial direction and $v_i$ the velocity vector, satisfying the relation

$$g_{ij}v^i v^j = 1 \quad (3)$$

Since the field is static, therefore

$$v^1 = v^2 = v^3 = 0 \text{ and } v^4 = \frac{1}{\sqrt{g_{44}}} \quad (4)$$

Thus we find that for the metric (1) under these conditions and for matter distributions with isotropic pressure the field equation (2) reduces to the following:

$$\frac{8\pi G}{c^4}p_r = e^{-\omega}\left[\frac{(\omega')^2}{4} + \frac{\omega'}{r} + \frac{\omega'\nu'}{2} + \frac{\nu'}{r}\right] + \frac{q^2}{r^4} \quad (5)$$

$$\frac{8\pi G}{c^4}p_\perp = e^{-\omega}\left[\frac{\omega''}{2} + \frac{\nu''}{2} + \frac{(\nu')^2}{4} + \frac{\omega'}{2r} + \frac{\nu'}{2r}\right] - \frac{q^2}{r^4} \quad (6)$$

$$\frac{8\pi G}{c^2}\rho = -e^{-\omega}\left[\omega'' + \frac{(\omega')^2}{4} + \frac{2\omega'}{r}\right] - \frac{q^2}{r^4} \quad (7)$$

where prime denotes differentiation with respect to r. From equations (5) and (6) we obtain following differential equation in $\omega$ and $\nu$.

$$\omega'' + \nu'' + \frac{(\nu')^2}{2} - \frac{(\omega')^2}{2} - \omega'\nu' - \frac{\omega'+\nu'}{r} = \left[\frac{16\pi G}{c^4}(p_\perp - p_r) + \frac{4q^2}{r^4}\right]e^\omega \quad (8)$$

Our task is to explore the solutions of equation (8) and obtain the fluid parameters $p_r$, $p_\perp$ and $\rho$ from equations (5), (6) and (7).

To solve the above equation we consider a seed solution as Pant et al. [12] and the electric intensity $E$ of the following form:

$$\frac{E^2}{C} = \frac{q^2}{Cr^4} = \frac{KCr^2}{B^2(1+Cr^2)^{\frac{12}{7}}} \quad (9)$$

Where $K$ is a positive constant defined as charge parameter. The electric intensity is so assumed that the model is physically significant and well behaved i.e. $E$ remains regular and positive throughout the sphere. In addition, $E$ vanishes at the center of the star and increases towards the boundary.

We also take,

$$\frac{8\pi G}{c^4}(p_\perp - p_r) = \Delta = \frac{2\alpha C^2 r^2}{B^2(1+Cr^2)^{\frac{12}{7}}} \quad (10)$$

where $\Delta$ is the anisotropy factor whose value is zero at the center and increases towards the boundary and $\alpha$ is a positive constant defined as anisotropy parameter for Esculpi et al. [31].

## 4. A New Class of Solution

In order to solve (8), we assume the following

$$e^{\omega/2} = B(1+Cr^2)^{-\frac{1}{7}}, x = Cr^2, \frac{d\nu}{dx} = \frac{2}{z}\frac{dz}{dx} \text{ and}$$
$$q^2 = \frac{K}{B^2C}x^3(1+x)^{\frac{-12}{7}} \quad (11)$$

On substituting the above in eq. (8), we get the following Riccati differential equation in $z$,

$$49(1+x)^2\frac{d^2z}{dx^2} + 14(1+x)\frac{dz}{dx} + \frac{12-49(K+\alpha)}{2} = 0 \quad (12)$$

which yields the following solution

$$e^{\frac{\nu}{2}} = \frac{\left\{1 + A(1+Cr^2)^{\frac{S}{7}}\right\}(1+Cr^2)^{\frac{5-S}{14}}}{B^2} \quad (13)$$

where A, B, C and K are arbitrary constants and

$$S = \sqrt{98(K+\alpha)+1} \quad (14)$$

$S$ is real for $K + \alpha \geq \frac{-1}{98} = -0.01$

The expressions for density, radial pressure and transversal pressure are given by

$$\frac{8\pi G\rho}{c^2} = \frac{C}{49B^2 f^{24}}\left(84 + 24Cr^2 - 49KCr^2\right) \quad (15)$$

$$\frac{8\pi G p_r}{c^4} = \frac{C}{49B^2 f^{24}\{1+Af^{2S}\}}$$
$$\times \begin{bmatrix} Cr^2[Af^{2S}(49K+26+10S)+(49K+26-10S)] \\ +14[Af^{2S}(3+S)+(3-S)] \end{bmatrix} \quad (16)$$

$$\frac{8\pi G p_\perp}{c^4} = \frac{C}{49B^2 f^{24}\{1+Af^{2S}\}}$$
$$\times \begin{bmatrix} Cr^2[Af^{2S}(49K+98\alpha+26+10S)+(49K+98\alpha+26-10S)] \\ +14[Af^{2S}(3+S)+(3-S)] \end{bmatrix} \quad (17)$$

$$\text{where }, f = (1+Cr^2)^{\frac{1}{14}} \quad (18)$$

## 5. Properties of the New Solution

The central values of pressure and density are given by

$$\left(\frac{8\pi G p_r}{c^4}\right)_{r=0} = \left(\frac{8\pi G p_\perp}{c^4}\right)_{r=0}$$
$$= \frac{2C}{7B^2(1+A)}[3(1+A)+S(A-1)] \quad (19)$$

$$\left(\frac{8\pi G\rho}{c^2}\right)_{r=0} = \frac{12C}{7B^2} \quad (20)$$

The central values of pressure and density will be non zero positive definite, if the following conditions will be satisfied.

$$A > \frac{S-3}{S+3} \text{ and } C > 0. \quad (21)$$

The pressure to density ratios are given by

$$\frac{p_r}{\rho c^2} = \frac{Cr^2[f^{2S}A(49K+26+10S)+(49K+26-10S)]+14[f^{2S}A(3+S)+(3-S)]}{(1+Af^{2S})[84+24Cr^2-49KCr^2]} \quad (22)$$

$$\frac{p_\perp}{\rho c^2} = \frac{Cr^2[f^{2S}A(49K+98\alpha+26+10S)+(49K+98\alpha+26-10S)]+14[f^{2S}A(3+S)+(3-S)]}{(1+Af^{2S})[84+24Cr^2-49KCr^2]} \quad (23)$$

Subjecting the condition that positive value of ratio of pressure-density and less than 1 at the centre i.e. $\frac{p_0}{\rho_0 c^2} \leq 1$ which leads to the following inequality,

$$\left[\frac{p_r}{\rho c^2}\right]_{r=0} = \left[\frac{p_\perp}{\rho c^2}\right]_{r=0} = \frac{3(1+A)+S(A-1)}{6(1+A)} = \frac{1}{2} - \frac{S(1-A)}{6(1+A)} \quad (24)$$

All the values of A which satisfy equation (21), will also lead to the condition $\frac{p_0}{\rho_0 c^2} \leq 1$.

Differentiating (16) with respect to r, we get

$$\frac{8\pi G}{c^4}\frac{dp_r}{dr} = \frac{C^2 r}{343 B^2 \{1+Af^{2S}\}^2 f^{38}} \times$$
$$\begin{bmatrix} -10\ Cr^2 \times [A^2 f^{4S}\{49K+26+10S\}+\{49K+26-10S\}+2Af^{2S}(49K+26-2S^2)] \\ +14\ [A^2 f^{4S}\{49K-46-14S\}+(49K-46+14S)+2Af^{2S}(49K-46+2S^2)] \end{bmatrix} \quad (25)$$

Thus it is found that extrema of $p_r$ occurs at the centre i.e.

$$p_r' = 0 \Rightarrow r = 0 \text{ and } \frac{8\pi G}{c^4}(p_r'')_{r=0} = -ve \quad (26)$$

Thus the expression of right hand side of equation (26) is negative for all values of A satisfying condition (21), showing thereby that the pressure ($p_r$) is maximum at the centre and monotonically decreasing.

Differentiating (17) with respect to r, we get

$$\frac{8\pi G}{c^4}\frac{dp_\perp}{dr} = \frac{C^2 r}{343 B^2 \{1+Af^{2S}\}^2 f^{38}} \times$$
$$\begin{bmatrix} -10Cr^2 \times [A^2 f^{4S}\{49K+98\alpha+26+10S\}+\{49K+98\alpha+26-10S\}+2Af^{2S}(49K+98\alpha+26-2S^2)] \\ +14[A^2 f^{4S}\{49K+98\alpha-46-14S\}+(49K+98\alpha-46+14S)+2Af^{2S}(49K+98\alpha-46+2S^2)] \end{bmatrix} \quad (27)$$

Thus it is found that extrema of $p_\perp$ occurs at the centre i.e.

$$p_\perp' = 0 \Rightarrow r = 0 \text{ and } \frac{8\pi G}{c^4}(p_\perp'')_{r=0} = -ve \quad (28)$$

Thus the expression of right hand side of equation (28) is negative for all values of A satisfying condition (21), showing thereby that the transversal pressure is maximum at the centre and monotonically decreasing.

Now differentiating equation (15) with respect to r we get

$$\frac{8\pi G}{c^2}\frac{d\rho}{dr} = \frac{C^2 r}{343 B^2 f^{38}}\left[-1680-686K-240Cr^2+490KCr^2\right] \quad (29)$$

Thus the extrema of $\rho$ occurs at the centre i.e.

$$\rho' = 0 \Rightarrow r = 0$$

$$\frac{8\pi G}{c^2}(\rho'')_{r=0} = -\frac{(1680+686K)C^2}{343B^2} \quad (30)$$

Thus, the expression of right hand side of (30) is negative showing thereby that the density $\rho$ is maximum at the centre and monotonically decreasing.

The square of adiabatic sound speed at the centre, $\frac{1}{c^2}\left(\frac{dp}{d\rho}\right)_{r=0}$, are given by

$$\frac{1}{c^2}\left(\frac{dp_r}{d\rho}\right)_{r=0} = \frac{(49K - 46)(1+A)^2 + 14S + 4AS^2 - 14SA^2}{(-120 - 49K)(1+A)^2} \quad < 1 \text{ and positive} \tag{31}$$

$$\frac{1}{c^2}\left(\frac{dp_\perp}{d\rho}\right)_{r=0} = \frac{(49K + 98\alpha - 46)(1+A)^2 + 14S + 4AS^2 - 14SA^2}{(-120 - 49K)(1+A)^2} \quad < 1 \text{ and + ve} \tag{32}$$

The causality condition is obeyed at the centre for all values of constants satisfying condition (21).

Due to cumbersome expressions the trend of pressure-density ratios and adiabatic sound speeds are studied analytically after applying the boundary conditions.

## 6. Boundary Conditions in Isotropic Coordinates

For exploring the boundary conditions, we use the principle that the metric coefficients $g_{ij}$ and their first derivatives $g_{ij,k}$ in interior solution (I) as well as in exterior solution (E) are continuous up to and on the boundary B. The continuity of metric coefficients $g_{ij}$ of $I$ and $B$ on the boundary is known first fundamental form. The continuity of derivatives of metric coefficients $g_{ij}$ of $I$ and $B$ on the boundary is known second fundamental form.

The exterior field of a spherically symmetric static charged fluid distribution is described by *Reissner-Nordstrom* metric given by

$$ds^2 = \left(1 - \frac{2GM}{c^2R} + \frac{q^2}{R^2}\right)c^2 dt^2 - \left(1 - \frac{2GM}{c^2R} + \frac{q^2}{R^2}\right)^{-1} dR^2 - R^2 d\theta^2 - R^2 \sin^2\theta\, d\varphi^2 \tag{33}$$

Where $M$ the mass of the ball as is determined by the external observer and $R$ is the radial coordinate of the exterior region.

Since *Reissner-Nordstrom* metric (33) is considered as the exterior solution, thus we shall arrive at the following conclusions by matching first and second fundamental forms in equations (1) and (33):

$$e^{\nu_b} = 1 - 2\frac{GM}{c^2 R_b} + \frac{q_b^2}{R_b^2} \tag{34}$$

$$R_b = r_b\, e^{\frac{\omega_b}{2}} \text{ and } q(r = r_b) = q_b \tag{35}$$

$$\frac{1}{2}\left(\omega' + \frac{2}{r}\right)_b r_b = \left(1 - 2\frac{GM}{c^2 R_b} + \frac{q_b^2}{R_b^2}\right)^{1/2} \tag{36}$$

$$\frac{1}{2}(\nu')_b\, r_b = \left(\frac{GM}{c^2 R_b} - \frac{q_b^2}{R_b^2}\right)\left(1 - 2\frac{GM}{c^2 R_b} + \frac{q_b^2}{R_b^2}\right)^{-1/2} \tag{37}$$

Equations (34) to (37) are four conditions, known as boundary conditions in isotropic coordinates.

Applying the boundary conditions from (34) to (37), we get the values of the arbitrary constants in terms of Reissner-Nordstrom parameter '$d$', Schwarzschild parameter $u = \frac{GM}{c^2 R_b}$ and radius of the star $R_b$.

$$C = \frac{7(1-d)}{(7d-5)r_b^2} > 0 \quad for\, u \le \frac{12}{49} + \frac{q_b^2}{2R_b^2} \tag{38}$$

$$A = \frac{7(u - q_b^2/R_b^2)(1 + Cr_b^2)^{(5-S)/14} - (5-S)Cr_b^2 d\,(1 + Cr_b^2)^{(-9-S)/14}}{(5-S)Cr_b^2 d\,(1 + Cr_b^2)^{(-9+S)/14} - 7(u - q_b^2/R_b^2)(1 + Cr_b^2)^{(5+S)/14}} \tag{39}$$

$$B = \sqrt{\frac{(1 + Cr_b^2)^{\frac{5-S}{14}} + A(1 + Cr_b^2)^{\frac{5+S}{14}}}{d}} \tag{40}$$

Where we define a new parameter called as Reissner-Nordstrom parameter '$d$' given by

$$d = \sqrt{1 - 2u + \frac{q_b^2}{R_b^2}} \tag{41}$$

whose value lies between $0.714 < d < 1$ for $Cr_b^2 > 0$.

Surface density is given by

$$\frac{8\pi G}{c^2}\rho_b R_b^2 = \frac{Cr_b^2[84 + 24Cr_b^2 - 49KCr_b^2]}{49(1 + Cr_b^2)^2} \tag{42}$$

Central red-shift is given by

$$Z_0 = \left[\frac{B^2}{1+A} - 1\right] \tag{43}$$

The surface red shift is given by

$$Z_b = [e^{-\frac{\nu_b}{2}} - 1] \tag{44}$$

The minimum period of rotation $P_{min}$ in millisecond of super dense star [32] is given by

$$P_{min} \cong 0.82 \left(\frac{R}{10km}\right)^{3/2} \left(\frac{M_\Theta}{M}\right)^{1/2} ms \qquad (45)$$

The precision of this formula (for subluminal equation of states (EOSs)) is within 5%. It is valid for both baryonic EOSs and for strange quark stars and other exotic stars.

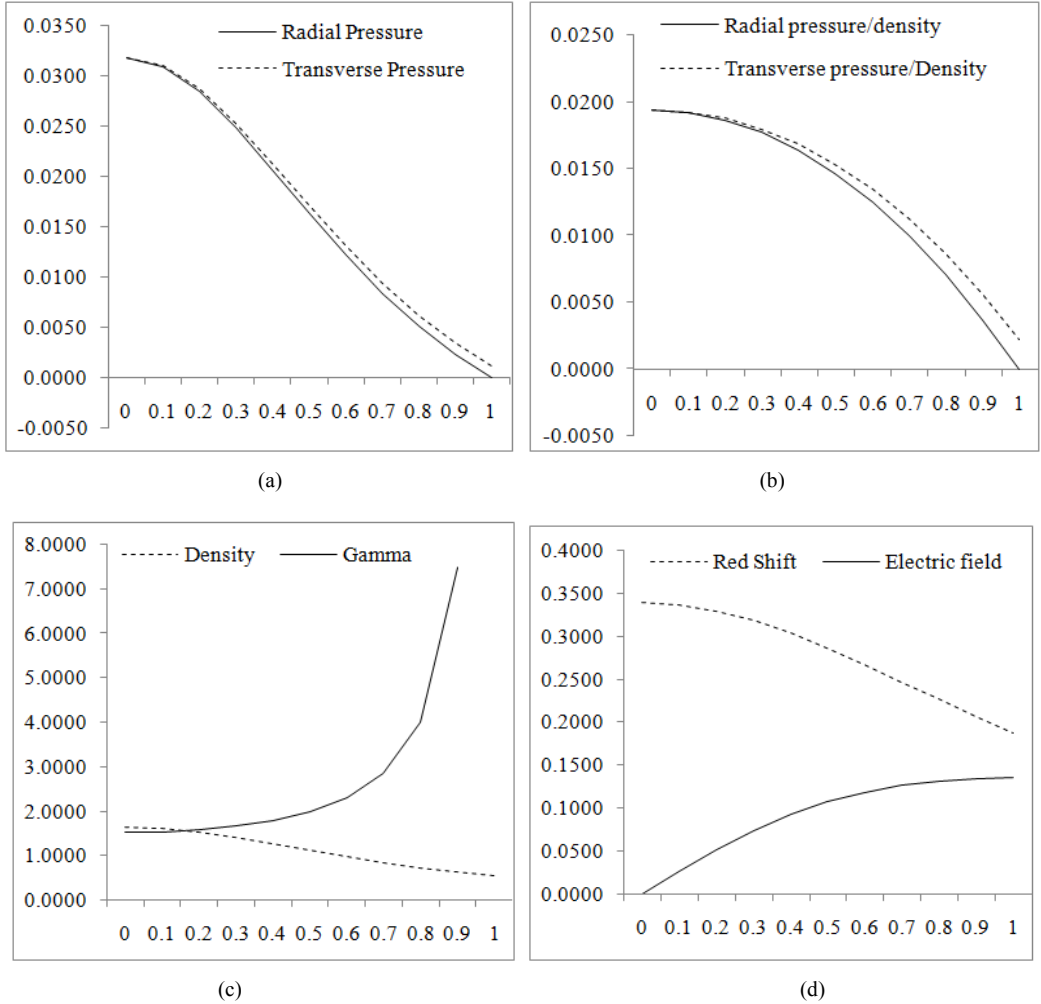

**Fig. 1.** The variations of $p_\perp$, $p_r$, $p_r/c^2\rho$, $p_\perp/c^2\rho$, $\rho, \gamma, E$ and $Z$ from centre to surface for $K = 0.062$ and $\alpha = 0.002$.

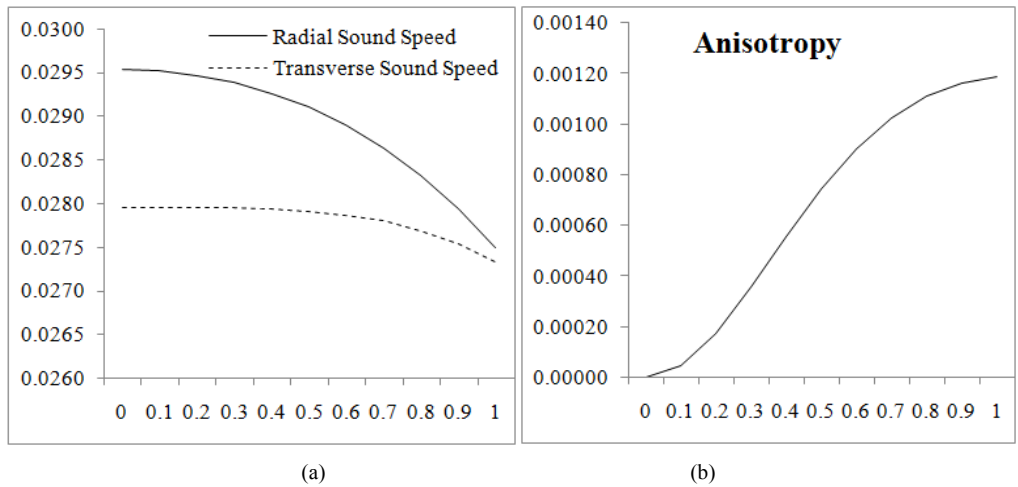

**Fig. 2.** The variation of square of sound speed ($dp_r/c^2 d\rho$ and $dp_\perp/c^2 d\rho$) and $\Delta$ from centre to surface for $K = 0.062$ and $\alpha = 0.002$.

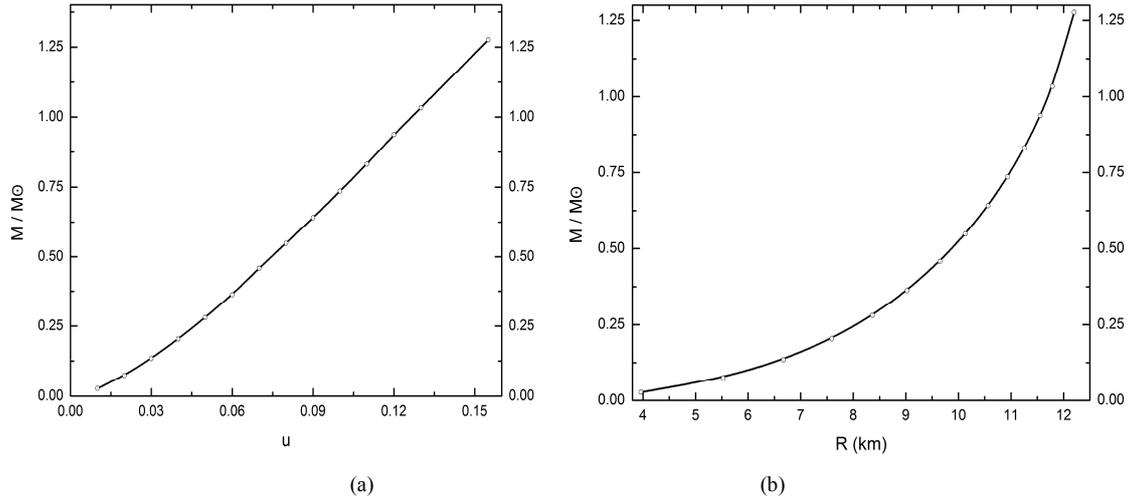

*Fig. 3. The variation of (a) mass of neutron star with u and (b) mass with radius*

*Table 1. Showing the variation of surface density, radius, maximum mass, surface red-shift, surface electric field and minimum period of rotation, with u for $K = 0.062$ and $\alpha = 0.002$.*

| $u$ | $8\pi G \rho\, r_b^2/c^2$ | $R\,(km)$ | $M/M_\odot$ | $z_b$ | $E_b r_b$ | $P_{min}\,(ms)$ |
|---|---|---|---|---|---|---|
| 0.010 | 0.06133106 | 3.961224 | 0.026693 | 0.01010 | 0.0089 | 1.251299 |
| 0.020 | 0.118626294 | 5.525944 | 0.074384 | 0.02041 | 0.0178 | 1.235051 |
| 0.030 | 0.171962686 | 6.673403 | 0.134582 | 0.03093 | 0.0266 | 1.218546 |
| 0.040 | 0.221417541 | 7.595131 | 0.20398 | 0.04167 | 0.0354 | 1.201774 |
| 0.050 | 0.267068535 | 8.366075 | 0.280517 | 0.05263 | 0.0441 | 1.184723 |
| 0.060 | 0.308993771 | 9.024959 | 0.362692 | 0.06383 | 0.0528 | 1.167381 |
| 0.070 | 0.350902111 | 9.647626 | 0.458186 | 0.07643 | 0.0623 | 1.147952 |
| 0.080 | 0.385259835 | 10.13695 | 0.548565 | 0.08814 | 0.0708 | 1.129953 |
| 0.090 | 0.416137584 | 10.56372 | 0.641455 | 0.10011 | 0.0794 | 1.111619 |
| 0.100 | 0.443616285 | 10.93527 | 0.736089 | 0.11235 | 0.0879 | 1.092931 |
| 0.110 | 0.467778005 | 11.25706 | 0.831761 | 0.12486 | 0.0963 | 1.073871 |
| 0.120 | 0.490624515 | 11.55851 | 0.937411 | 0.13895 | 0.1056 | 1.052451 |
| 0.130 | 0.508094671 | 11.78825 | 1.033145 | 0.15207 | 0.1140 | 1.032541 |
| 0.155 | 0.53946717 | 12.20438 | 1.276229 | 0.18765 | 0.1358 | 0.97864 |

## 7. Discussions and Conclusions

It has been observed that from Fig.1(a), (b), (c), (d) and Fig.2(a), the physical parameters $p_r, p_\perp, p_r/c^2\rho, p_\perp/c^2\rho, dp_r/c^2 d\rho, dp_\perp/c^2 d\rho, z$ are positive at the centre and within the limit of realistic state equation and monotonically decreasing while the parameters $\gamma, \Delta, E$ are increasing as seen from Fig.1(c), (d) and Fig.2(b) for $0.155 \geq u > 0$. If we increase the value of $\alpha$ above 0.002 while $K=0.062$ is kept constant or $K$ increases above 0.062 with $\alpha = 0.002$, the causality condition is obeyed throughout within the ball but the trend of adiabatic sound speed (transversal) is erratic. Thus, the solution is well behaved for all values of $u$ satisfying the inequality $0 < u \leq 0.155$ for K = 0.062 and α = 0.002. For $K + \alpha > 0.081$, the solution is not well behaved. With the increase in the value of anisotropy parameter $\alpha$, charge parameter $K$ decreases, therefore Schwarzschild parameter '$u$' decreases, hence the mass of the star decreases. With α=0 we recover the isotropic model of Pant et al. [33] and for $\alpha = 0, K = 0$, our solution reduce to Pant et al [12]. As we see from Table 1, it is shown that the variation of surface density, radius, maximum mass, surface red-shift, surface electric field and minimum period of rotation, with $u$ for $K = 0.062$ and $\alpha = 0.002$. From the Fig. 3 (a) and Table 1, we can observe that, by increasing the Schwarzschild parameter $u$, the mass of neutron star also increases. Similarly Fig. 3 (b) shows the variation of mass with radius obtained from our solution.

We now here present a model of anisotropic neutron star based on the particular solution discussed above by assuming surface density; $\rho_b = 2 \times 10^{14} g/cc$ corresponding to $K_{max} = 0.062, \alpha = 0.002, u = 0.155$, resulting a well behaved solution having maximum mass $M = 1.276\, M_\odot$ and radius $R = 12.204\, km$ with minimum period of rotation $0.979\, ms$. For isotropy, $\alpha = 0$ and taking $K = 0.062$, the super dense star has maximum mass $1.412 M_\odot$ and radius $12.33\, km$ which is a well behaved solution. Therefore, in the presence of anisotropy, maximum mass of the super dense star reduces. This is because of the fact that in pressure isotropy there is only one component of pressure which is along radial direction that can support more masses. However, in a pressure anisotropic star some of the pressure gives rise to tangential pressure and hence the effective pressure in radial direction reduces and thus mass that can support by the radial pressure also decreases.